 \documentclass[letterpaper,journal]{IEEEtran}

\ifCLASSINFOpdf
 
\else
 
\fi

\usepackage{graphicx}
\usepackage{subfig}
\usepackage{cite}
\usepackage{color}
\usepackage{amsmath}
\usepackage{amssymb}
\usepackage[font={small}]{caption}
\usepackage{algorithm}
\usepackage[noend]{algpseudocode}
\usepackage{forest}
\usepackage{tikz-qtree}
\usepackage{array}
\usepackage{longtable}
\usepackage{epsfig} 

\begin{document}
% \onecolumn
% \title{Block SHM: Blockchain-based approach for Structural Health Monitoring of Bridges using Novelty Index } 
\title{BIONIB: Blockchain-based IoT using Novelty Index in Bridge Health Monitoring}
\author{Divija Swetha Gadiraju, Ryan McMaster, Saeed Eftekhar Azam, and Deepak Khazanchi
\thanks{D. S. Gadiraju,  Ryan McMaster, and D. Khazanchi are with the University of Nebraska at Omaha. S.E. Azam is with the University of New Hampshire.\ {This work is partially supported by contracts W912HZ21C0060 and W912HZ23C0005, U.S. Army Corps of Engineers, Engineering Research and Development Center (ERDC).}% email: \{dgadiraju,smuthiah,khazanchi\}@unomaha.edu\. 
\\ Email:  \{dgadiraju,khazanchi\}@unomaha.edu, saeed.eftekharazam@unh.edu \
}}%, and iHub Anubhuti-IIITD Foundation.}}

\if 0
\IEEEauthorblockA{\IEEEauthorrefmark{1}Purdue University, USA} 
\IEEEauthorblockA{\IEEEauthorrefmark{2} International Institute of Information Technology, Hyderabad, India } 
%\IEEEauthorblockA{\IEEEauthorrefmark{3} King Abdulaziz University, Saudi Arabia } 
Email: \IEEEauthorrefmark{1} \{dgadiraj, vaneet\}@purdue.edu \
\IEEEauthorrefmark{2} email\ 
%\IEEEauthorrefmark{3} analahmadi@kau.edu.sa

\fi 

\newcommand{\ourmethod}{BIONIB}%{this work}

% \onecolumn

\maketitle

\begin{abstract}

Bridge health monitoring becomes crucial with the deployment of IoT sensors. The challenge lies in securely storing vast amounts of data and extracting useful information to promptly identify unhealthy bridge conditions. To address this challenge, we propose \ourmethod, wherein real-time IoT data is stored on the blockchain for monitoring bridges. One of the emerging blockchains, EOSIO is used because of its exceptional scaling capabilities for monitoring the health of bridges. The approach involves collecting data from IoT sensors and using an unsupervised machine learning-based technique called the Novelty Index (NI) to observe meaningful patterns in the data. Smart contracts of EOSIO are used in implementation because of their efficiency, security, and programmability, making them well-suited for handling complex transactions and automating processes within decentralized applications. \ourmethod\ provides secure storage benefits of blockchain, as well as useful predictions based on the NI. Performance analysis uses real-time data collected from IoT sensors at the bridge in healthy and unhealthy states. The data is collected with extensive experimentation with different loads, climatic conditions, and the health of the bridge. The performance of \ourmethod\ under varying numbers of sensors and various numbers of participating blockchain nodes is observed. We observe a tradeoff between throughput, latency, and computational resources. Storage efficiency can be increased by manifolds with a slight increase in latency caused by NI calculation. As latency is not a significant concern in bridge health applications, the results demonstrate that \ourmethod\ has high throughput, parallel processing, and high security while efficiently scaled. 

\end{abstract}

% % Note that keywords are not normally used for peerreview papers.
% \begin{IEEEkeywords}
% {Blockchain, IoT, Bridge Health, Structural Health Monitoring.}
% \end{IEEEkeywords}

% \keywords{Deep reinforcement learning, neural networks, Bridges, structural health monitoring}

\section{Introduction}
Bridge health monitoring is a crucial aspect of infrastructure management, especially with the integration of Internet of Things (IoT) technologies\cite{shmliterature}. IoT sensors installed on bridges allow for the continuous and real-time collection of data, providing valuable insights into the structural health and performance of these vital components of transportation networks \cite{DRL_sensor, IoTbridgereinforcement}. However, the large amounts of data generated by these sensors pose a significant challenge in terms of secure storage and meaningful analysis. Traditional methods of data storage and analysis may not be effective in handling this vast amount of information \cite{sc_civil_blockchain}. Moreover, timely and accurate observations are crucial to ensure the safety and longevity of bridges. Therefore, it is essential to explore innovative approaches that not only tackle the challenges of data storage but also enhance the ability to derive actionable insights from the collected data.
Blockchain technology provides a promising solution to the problem of secure storage. Due to its decentralized and tamper-resistant nature, blockchain is an ideal choice for ensuring the integrity and security of the vast datasets generated by IoT sensors in bridge health monitoring \cite{BCapplications_survey,bc_iothybrid}. The use of smart contracts further amplifies the capabilities of blockchain by enabling programmable and automated actions based on predefined conditions \cite{arcbridge_survey_deployment,unsupervised_blockchain}.

We propose Blockchain-based IoT using Novelty
Index in Bridge health monitoring,  \ourmethod, aims to transform bridge health monitoring by utilizing the EOSIO blockchain. The main objective is to create a seamless integration of IoT sensor data, blockchain technology, and advanced data analysis techniques to offer an efficient and robust solution. Specifically, we focus on using an unsupervised learning-based technique known as the novelty index (NI) to identify meaningful patterns in the IoT data \cite{NIunsupervised,NIevaluating}. This approach will enable the rapid detection of unhealthy conditions in bridges.
Our proposed methodology involves storing real-time data from IoT devices on the blockchain and using smart contracts to extract valuable predictions based on the NI. This not only enhances secure data storage but also significantly improves the effectiveness of bridge health monitoring systems. In the following sections, we will provide a detailed explanation of our approach, including its implementation using the EOSIO blockchain platform \cite{eosio_security}. EOSIO is a high-performance platform with excellent scalability and high throughput capabilities \cite{EOSIOsurvey}. We will present the results of performance analyses based on real-world data collected from IoT sensors deployed on both healthy and unhealthy bridges. These results demonstrate the flexibility and advantages of our approach across various scenarios, highlighting its potential to advance the field of structural health management (SHM), especially bridge health monitoring.

In \ourmethod\, we address the challenge of secure storage of IoT data by using the EOSIO blockchain with smart contracts. The real-time data collected from the sensors is used in our experimentation. In addition, bridge health can be monitored from IoT data with a NI proposed in \cite{NIunsupervised,NIevaluating}. The storage efficiency is improved by using the NI for data storage rather than using IoT data on the blockchain. The
performance of \ourmethod\ is evaluated for varying numbers of sensor data. We observe a tradeoff between throughput,
latency, and computational resources. The CPU resources are used fairly consistently because of EOSIO's parallel processing capabilities \cite{EOSIOsurvey}. However, with more number sensors, the execution time is higher at the smart contract and the throughput is also higher as more blocks are confirmed on the blockchain. The results demonstrate a detailed analysis of various scenarios where this \ourmethod\ proves to be advantageous.

The remainder of this article is organized as follows. Section II discussed the related work. The network architecture is presented in Section III. Data collection and analysis are discussed in Section IV. The proposed algorithm is presented in Section V.  Section VI shows the performance analysis. Section VI concludes the article with concluding remarks and future research direction.

\section{Related Work}
Blockchains have recently become popular in IoT, healthcare, smart-grid, supply-chain, and many other fields for immutability, secure storage, and decentralization \cite{abou2019blockchain}. Many studies have been conducted concentrating on the applications of Blockchains for specific IoT sensors and tasks \cite{wu2019comprehensive}. In this section, we present a review of the literature related to our research.

\subsection{IoT based Blockchain}
In \cite{BCapplications_survey}, a survey is presented showing various applications of blockchain in IoT. The survey \cite{BCapplications_survey} also focuses on the challenges posed by the current centralized IoT models, and recent advances made both in industry and research to solve these challenges for effectively using blockchains to provide a decentralized, secure medium for the IoT. The use of blockchain in Industry 4.0 is presented in \cite{ind4.0blockchain}. The merits and demerits of traditional security solutions in blockchains are discussed in comparison to their countermeasures and a comparison is provided \cite{ind4.0blockchain}.  Various consensus protocols like sharding \cite{gadiraju2020secure,cai2021sharding} are used in IoT-based blockchains. Recent works focus on using various deep learning and scaling algorithms for blockchain for IoT applications \cite{khan2020iot,10038442,aljuhani2023deep}. These protocols demonstrate the usability of the blockchain consensus algorithms in the domain of IoT.

\subsection{SHM}
Bridge health monitoring by analyzing the survey data for bridges is presented in \cite{du2022parameterized}. 
In\cite{arcbridge_survey_deployment}, the recent progress of the SHM technology is reviewed for long-span bridges. In addition to the data analysis and condition assessment including techniques on modal identification, methods on signal processing, and damage identification, a case study about a SHM system of a long-span arch bridge is discussed.
The authors in \cite{shmliterature} conducted a detailed literature review of recent applications of smartphones, UAVs, cameras, and robotic sensors for structural condition monitoring and maintenance. The work in \cite{TIMESERIESstructural}  employs a nondestructive evaluation test with statistical confidence and uses time-series analysis instead of frequency-domain monitoring. 

The study in \cite{NIevaluating,NIunsupervised} developed a damage detection tool using unsupervised machine learning with data from tests of a full-scale bridge deck mock-up. The NI was validated against field data from a full-scale mock-up bridge. The NI will be explained in the subsequent sections. %The authors explored the use of Singular Value Decomposition (SVD) and Independent Component Analysis (ICA) as damage sensitive features. Novelty indices were developed using Left Singular Vectors (LSVs) and Independent Component Modes (ICMs). The method has the potential to be a robust damage detection tool, but further investigation is required to determine the method's application for detecting damage types other than the one studied herein due to the statistical nature of SVD.

\subsection{Blockchain in SHM}
The work in \cite{bc_iothybrid} combines IoT and blockchain-based smart contracts for SHM of underground structures to create an efficient, scalable, and secure network that enhances operational safety. The authors in \cite{definingbc} propose a conceptual framework to integrate vibration-based methods and blockchain for more reliable and efficient structural damage detection. The system in \cite{sc_civil_blockchain} is a blockchain network for SHM, with smart contracts for health monitoring on the Ethereum private chain for verifying authority, detecting structural damage, generating alerts, securing data, and allowing for traceability queries. In \cite{unsupervised_blockchain}, sensors record monitoring information such as pressure points, temperature, and pre-tension force. This data is transmitted to a blockchain platform that classifies transaction criticality and securely stores  the information.

\subsubsection{Why EOSIO for IoT SHM?}
In \cite{EOSIOxblock}, the authors collect and process the up-to-date on-chain data from EOSIO.  EOSIO achieves a significant improvement in performance when compared to Bitcoin and Ethereum \cite{EOSIOxblock} since it generates more blocks and hence provides more transactions per second. In \cite{EOSIO_scalability_revisiting}, the authors analyze the network traffic of three major high-scalability blockchains (EOSIO, Tezos, and XRP Ledger) for seven months and find that a small portion of transactions are used for value transfer purposes. About 96\% of the transactions on EOSIO were triggered by the airdrop of a valueless token \cite{EOSIO_scalability_revisiting}.

The authors in \cite{eosio_security} propose EOSAFE which is a static analysis framework that detects vulnerabilities in EOSIO smart contracts at the bytecode level and analyze EOSIO smart contracts against attacks.
In \cite{EOSIOdecentralization}, a systematic analysis is conducted on the decentralization of DPoS with data from blocks in EOSIO, the first DPoS-based blockchain system, and characterize the decentralization into two phases, the block producer election, and the block production. 
The method in \cite{EOSIOchronoeos} can record the events that occur in multiple industrial robotic arms by deploying a Smart Contract in the EOSIO blockchain.
The authors in \cite{EOSIOsurvey} analyzed attacks on EOSIO and found vulnerabilities in its components and outlined effective mitigations and best practices for researchers. 

Our selection of EOSIO was based on the presence of a well-supported research community and a rigorous security analysis. Additionally, EOSIO offers exceptional scalability making it suitable for IOT based SHM.

\subsubsection{Comparison with central storage}
The IoT data is stored on a central server traditionally. Blockchain is particularly useful for better decentralization, security, and tamper resistance properties. Storage efficiency refers to the effective storage and management of data over time. EOSIO \cite{EOSIOsurvey} is one of the emerging blockchains that offers scalability and flexible operability for our implementation. In EOSIO, the storage efficiency is impacted by factors such as the number of nodes, smart contracts, and additional metadata. On the other hand, centralized storage is not dependent on the number of nodes. EOSIO Blockchain's storage efficiency scales linearly with the number of nodes, making it more advantageous. The time it takes to retrieve data in a blockchain network, known as data retrieval time, is influenced by the number of nodes, and the time data takes to propagate between nodes. The scalability of a blockchain network depends on its ability to handle a growing number of nodes and transactions while maintaining high performance. In contrast, the scalability of centralized storage is limited by the capacity of the central server. EOSIO Blockchain is highly scalable and can efficiently handle a larger network size, particularly when it comes to transaction processing. Blockchain technology is highly secure and resistant to malicious activities due to its decentralized and immutable nature. It offers superior protection against security breaches and tampering compared to centralized systems.

\begin{figure}
    \centering
    \includegraphics[width=0.9\linewidth]{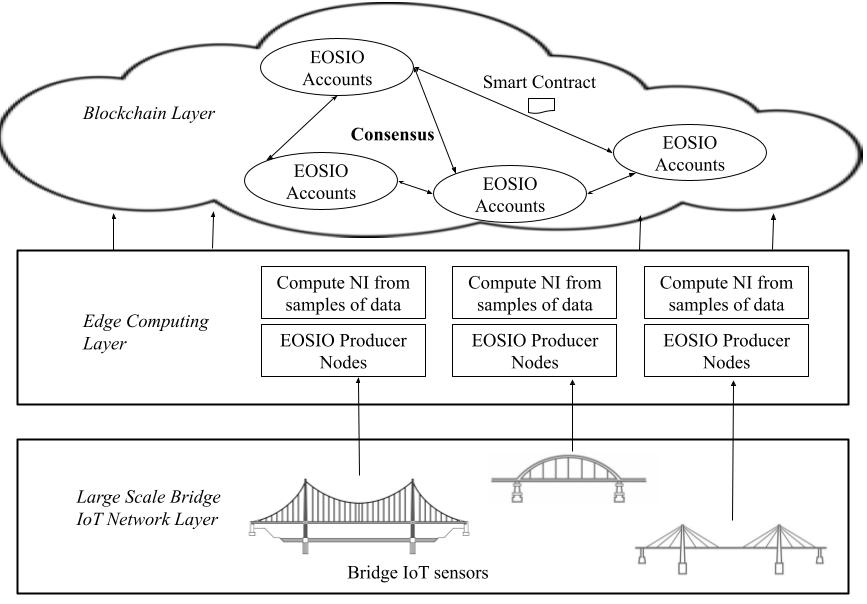}
    \caption{Network Architecture of Blockchain integrated IoT Bridge Network}
    \label{fig:sys}
\end{figure}

\section{Network Architecture}
In this section, we will explain the EOSIO  consensus used for our implementation. Later, we will discuss the EOSIO protocol with smart contracts.
Fig. \ref{fig:sys} shows the network architecture of Bridge monitoring with sensors and Blockchain. Consider $N$ number of block producer nodes and $k$ accounts created. The Blockchain Layer has the EOSIO Accounts and the Block Producer nodes which confirm the input transactions into the blocks on the blockchain. For better understanding, EOSIO Producer nodes are shown in the Edge Computing Layer but they are also a part of the blockchain layer. Let $s$ be the number of strain sensors collecting the data on every bridge. Each bridge sensor transmits its data to the edge computing layer which consists of the block producer nodes who calculate the NI from the received sensor data. These Novelty Indices are given to the EOSIO accounts who initiate a smart contract. The deployed smart contracts then output the actions, which are the transactions (containing novelty indices), which after EOSIO consensus are added to the blockchain.

\subsection{EOSIO Consensus}
EOSIO \cite{EOSwhite},\cite{decentralization} is a blockchain protocol designed to empower the development and deployment of decentralized applications (DApps) with a focus on scalability, flexibility, and user-friendliness. EOSIO is a blockchain platform that uses a Delegated Proof-of-Stake (DPoS) consensus mechanism. The consensus mechanism has two distinct components: Producer voting and scheduling, which is executed by the DPoS layer 2, and Block production and validation, which is carried out by the native consensus layer 1. Additionally, EOSIO uses asynchronous Byzantine Fault Tolerance (aBFT) in Layer 1 to ensure safety and consistency in the validation process. In Delegated Proof-of-Stake (DPoS), a small group of elected nodes, called block producers, are responsible for verifying transactions and adding blocks to the blockchain. Compared to the traditional Proof-of-Work (PoW) consensus mechanism, this design improves scalability and transaction throughput. EOSIO uses parallel processing, allowing multiple transactions to be processed at the same time, resulting in increased transaction speed and efficiency. It is ideal for applications that require high performance and low latency. Additionally, EOSIO introduces a resource model that includes bandwidth, computation, and storage. 

Layer 1 uses aBFT (Asynchronous Byzantine Fault Tolerance), an advanced consensus algorithm, to validate blocks produced by elected producers and record them on the blockchain permanently. The layer ensures that each block is signed by the corresponding producer by utilizing the proposed schedule of producers from Layer 2. To ensure byzantine fault tolerance, this layer follows a two-stage block confirmation process, where each block must be confirmed twice by a two-thirds supermajority of producers from the currently scheduled set. The layer proposes a last irreversible block (LIB) in the first stage, and the proposed LIB is confirmed as final in the second stage, making the block irreversible. Additionally, any changes made to the producer schedule are notified at the beginning of every schedule round. 
The EOSIO consensus mechanism attains algorithmic finality, a departure from the probabilistic finality characteristic of Proof of Work models. Active producers on the EOSIO blockchain use signatures to validate each block. The schedule determines authorized signers for each block at specific time slots. Smart contracts can alter the schedule, but changes are implemented only after the block has undergone two stages of confirmation.  EOSIO smart contracts allow us to define specific rules and conditions for transactions. 
In the proposed approach for bridge health monitoring, EOSIO blockchain integration involves deploying smart contracts to handle the secure storage of real-time IoT data and execute predefined logic for data analysis. This incorporation enhances the overall security, transparency, and efficiency of the bridge health monitoring system. 

\begin{figure*}[]
  \centering
  \begin{minipage}[b]{0.5\linewidth}
    \centering
    \includegraphics[width=\linewidth]{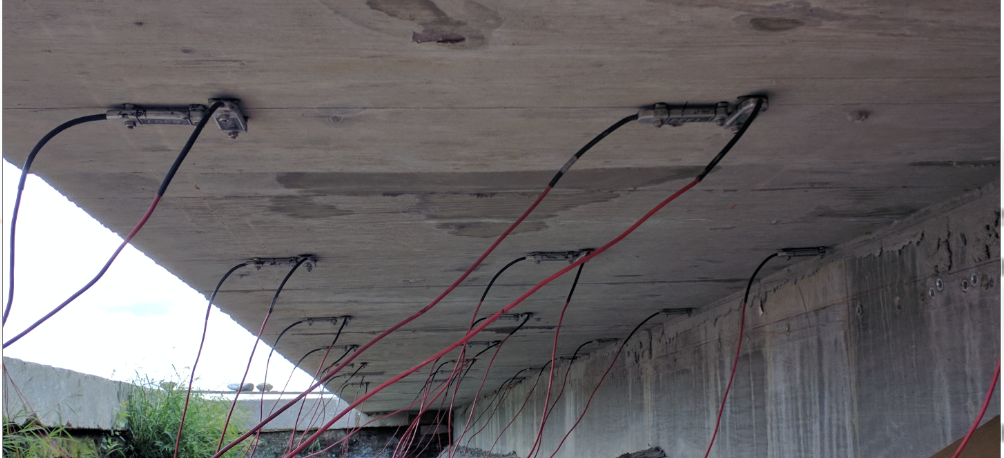}
    \caption {An image of sensors placed on the bottom of the bridge deck.}
    \label{fig:bridge1}
  \end{minipage}
  \hfill
  \begin{minipage}[b]{0.45\linewidth}
    \centering
    \includegraphics[width=\linewidth]{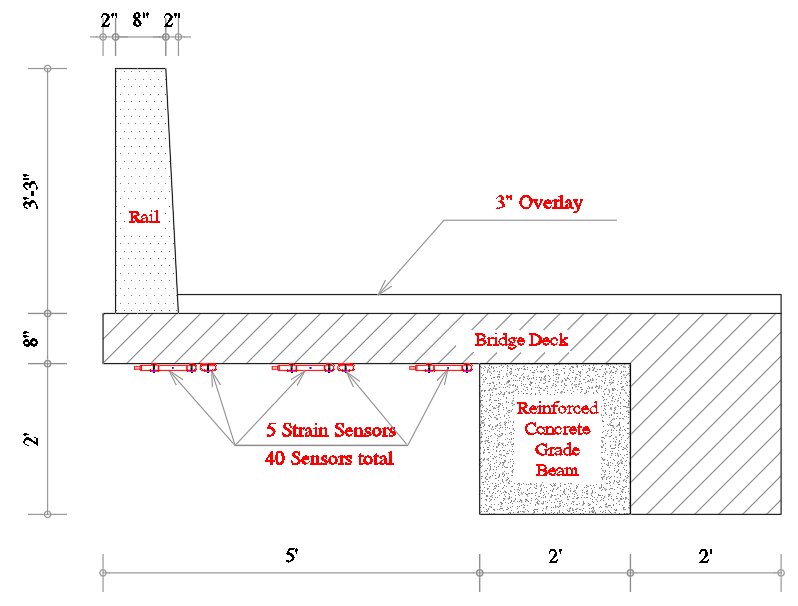}
    \caption {A illustration of half of the bridge with deployed sensors. The other half follows the same pattern.}
    \label{fig:bridge2}
  \end{minipage}
    \label{fig:sensors}
\end{figure*}

\section{Data collection and Analysis}

% \begin{figure}
%     \centering
%     \includegraphics[width= 0.9\linewidth]{sysmodel.png}
%     \caption{}
%     \label{fig:sysmodel}
% \end{figure}

\subsection{Collection of Data}

Field tests were conducted at the Midwest Roadside Safety Facility (MwRSF) to evaluate the barrier performance during vehicle impacts on a bridge deck mock-up. The authors in \cite{NIevaluating} measured the bridge's response to vehicle loads at different speeds and under various levels of damage before and after the crash test.
Fig. \ref{fig:bridge1} and Fig. \ref{fig:bridge2} show a cantilevered bridge deck with a concrete barrier at the end. It's composed of a grade beam, deck, barrier, and overlay. The mock-up mimics a real bridge. The grade beam replicates a girder, and the design matches realistic forces. The overhang and barrier meet AASHTO's and NCHRP's design specifications. The bridge deck had 40 BDI strain transducers placed across its width at eight locations. Each section was equipped with 2 longitudinal and 3 transverse strain sensors, out of which 24 sensors were used for the study. Fig. \ref{fig:bridge2} illustrates the locations of the sensors on one half of the bridge. The other half of the bridge follows the same sensor placement. The proposed damage detection method for a bridge was tested using a pickup truck and a dump truck. The bridge was tested in different conditions, with the pickup truck traveling at different speeds, while the dump truck moved slowly, both empty and loaded. The initial tests were carried out on the bridge in its original state, followed by tests with different levels of damage. Strain data were collected for each test, and the damage scenarios were designed to mimic real-world situations.

\begin{figure}
    \centering
    \includegraphics[width = 0.9\linewidth]{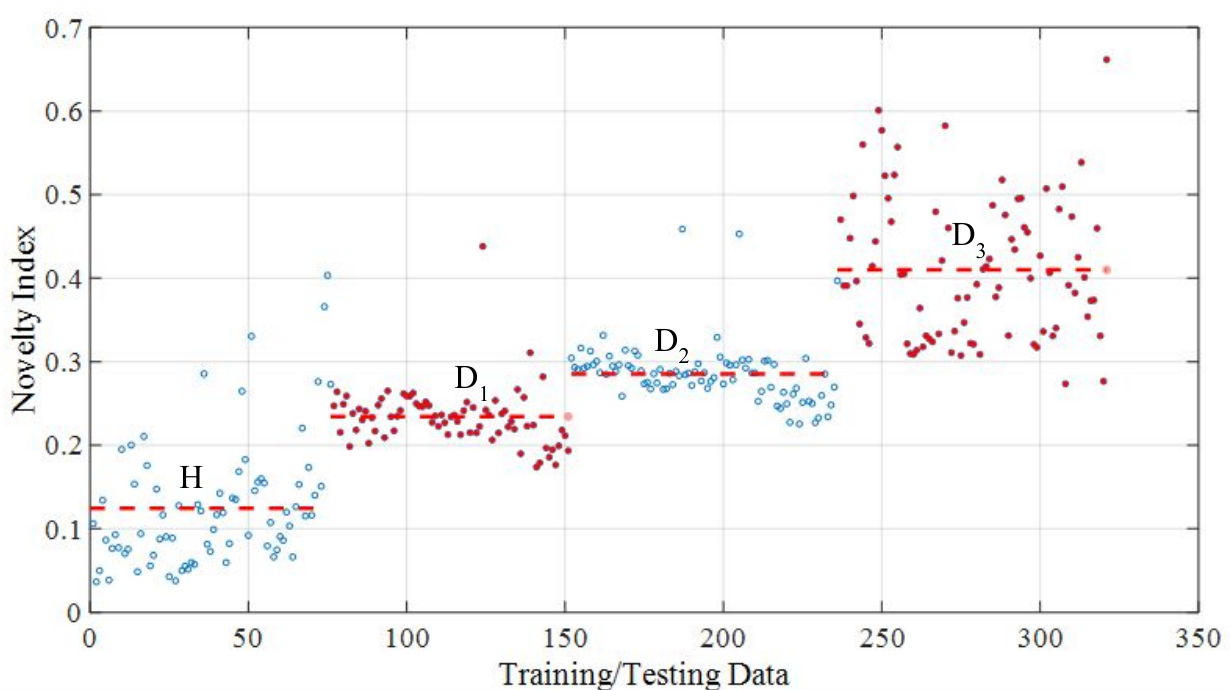}
    \caption{NI plot for collected data where the data points are NIs. Three damage levels along with a healthy state are shown.}
    \label{fig:NI}
\end{figure}

\subsection{Novelty Index calculation}
In this work we use NI calculations based on \cite{NIevaluating, NIunsupervised}. To detect damage, features are extracted from response data. Linear modal properties like resonant frequencies and mode shape curvatures are used to fit a linear, dynamic model to both healthy and damaged structure data. However, these features have limitations, such as being unable to detect nonlinear changes in system response and being influenced by environmental conditions. Recent research has focused on developing damage-sensitive features such as novelty analyses using Singular Value Decomposition (SVD), and Independent Component Analysis (ICA). These features transform measured, nonlinear response data into low-dimensional features, serving as damage indicators and input parameters for novelty indices. SVD and ICA offer computational simplicity compared to traditional vibration modes. It is important to note that strain transducers were used for live load response measurement and Operational Modal Analysis (OMA) was not utilized. SVD can be applied to both linear and nonlinear systems and is robust to measurement noise, while ICA is a potential alternative known for data reduction and handling measurement noise effectively.
Proper Orthogonal Decomposition (POD) generates orthonormal vectors ordered by importance. It has an intuitive link with PCA, which transforms correlated variables into principal components. In PCA used for POD, the first POM captures the most variability in the data. SVD and Karhunen–Lo`eve Decomposition (KLD) are other POD techniques. Numerically, PCA and SVD provide similar results. SVD was chosen to develop POMs in this study, but PCA could also be used.
Then we use datasets from bridge mock-ups, where $ U = [ u_1, u _2, \cdot, u_{ns} ] $ represents snapshot matrices of vehicle passage data. The SVD of these matrices allows for the extraction of damage features, represented as $ U = L\Sigma R$, where $L$, $\Sigma$, and $R$ are orthonormal matrices. The first Proper Orthogonal Mode (POM), $\phi_1$, contains information relevant to the presence of structural damage and is used as one of the damage features.

This work in \cite{NIevaluating} proposes a detection framework to differentiate between healthy and damaged states in bridge response data from vehicle passages. Fig. \ref{fig:NI} shows how NI summarizes the state of the bridge based on bridge sensor data samples. The experiment is conducted initially on a healthy bridge and then on unhealthy bridge. The scatter plot shows the clustering of NIs calculated from the strain transducers. In Fig. \ref{fig:NI}, $H$ represents the healthy bridge state without any damages. The damage levels $D_1$, $D_2$, and $D_3$ are the bridge condition after introducing damage to the structure. The damage levels are in the increasing order of their severity. The concrete bridge deck is saw-cut to introduce the three damage levels as described in \cite{NIunsupervised}. The damage level $D_3$ is a complete saw cut deck damage and hence the NI is closer to 1. However, even at $D_3$ the bridge has not completely failed and we could conduct the experiments with load on the bridge.  NI is calculated first, and the feature extraction is performed using SVD. Then healthy feature vectors, denoted by $h$ are stored. Next, we calculate the healthy feature vector mean for POM
\begin{equation}
    \Bar{\chi} = \frac{\Sigma_{i=1}^{h}\chi^i}{h}.
\end{equation}
 
After this, we move to calculate $\mathcal{N}^i$ for each data set with Euclidean norm
 \begin{equation}
     \mathcal{N}^i = || \chi^i - \Bar{\chi} ||.
 \end{equation}
The detailed procedure of the usage of NI in \ourmethod\ is explained in the subsequent sections.

\section{Proposed Approach: BIONIB}

% \begin{figure}
%     \centering
%     \includegraphics{}
%     \caption{Caption}
%     \label{fig:enter-label}
% \end{figure}

\subsection{EOSIO Integration with IoT data}
\ourmethod\ utilizes EOSIO, a secure, transparent, and decentralized platform to store and access NI data from IoT sensors. This ensures data integrity and immutability while facilitating efficient data management and analysis.  Bridge IoT sensors collect measurement data, such as strain from bridges. This data is transmitted to a blockchain system for processing and analysis. The collected data undergoes processing to extract relevant features and calculate an NI. Feature extraction technique, SVD is used to extract information from the sensor data. The NI is calculated based on the extracted features to assess the deviation or abnormality of the bridge's condition from a baseline. A smart contract is developed using the EOSIO Contract Development Toolkit (EOSIO.CDT) \cite{eosio-github}. The smart contract includes actions and data structures necessary for adding and retrieving NI data to and from the blockchain. The compiled smart contract is deployed to the EOSIO blockchain using a deployment transaction. This transaction specifies the account deploying the contract. Once deployed, users can interact with the smart contract by invoking its actions.
An action within the smart contract is created to add the NI data to the blockchain. This action includes parameters such as the bridge identifier and the calculated NI. The smart contracts are invoked and the action adds the NI data to the blockchain by sending a transaction. The transaction is processed by EOSIO nodes in the network, validated, and included in a block by block producers through the consensus mechanism. Once included in a block and confirmed by subsequent blocks, the transaction becomes immutable and part of the blockchain's permanent record. Users, applications, or external systems can query the blockchain to retrieve NI data stored by the smart contract. EOSIO APIs enable easy access and transparency to stored data. Nodeos handles the blockchain data persistence layer, peer-to-peer networking, and contract code scheduling. The command 'cleos' is a command line tool that interfaces with the REST APIs exposed by nodeos.

% \subsubsection{Comparision of with and without smart contract}
 Using a smart contract has several advantages such as structured data management, automation, decentralized trust, and efficiency. However, it also comes with some drawbacks like development complexity, execution costs, and maintenance requirements. On the other hand, without a smart contract, users may have to rely on centralized systems and experience higher costs and processing times \cite{EOSIOsurvey}.

 \begin{figure}
    \centering
    \includegraphics[width=0.9\linewidth]{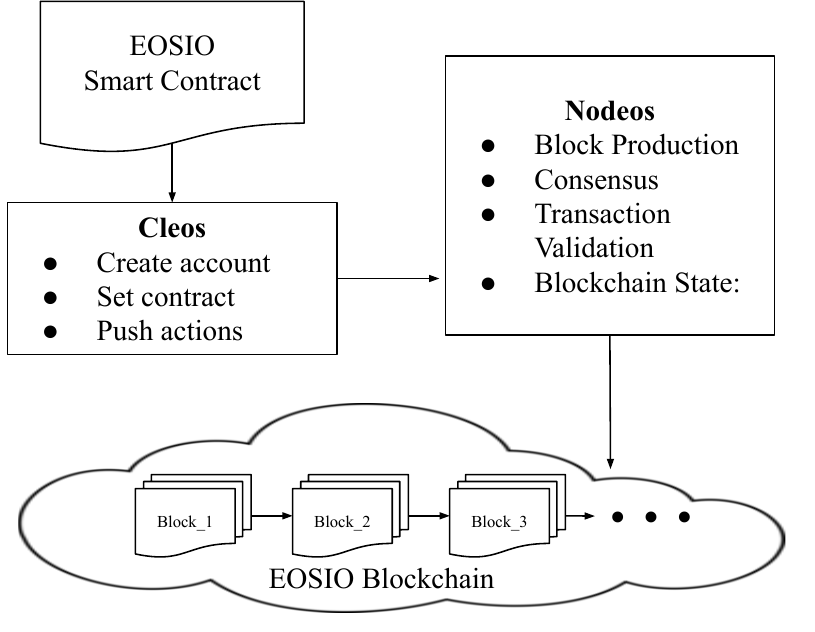}
    \caption{EOSIO smart contract workflow}
    \label{fig:sc-workflow}
\end{figure}

\begin{algorithm}
\caption{\ourmethod}
\label{alg:iot_blockchain}
\begin{algorithmic}[1]
 \State \textbf{Input:} IoT bridge data
 \State Measure IoT sensor response data
 \State Feature extraction using SVD
 \State Store all healthy feature vectors $h$
 \State Calculate healthy feature vector mean for Proper Orthogonal Modes (POM) 
 $$
\Bar{\chi} = \frac{\Sigma_{i=1}^{h}\chi^i}{h} 
 $$
 \State Calculate $\mathcal{N}^i$ for each data set with Euclidean norm
 $$
\mathcal{N}^i = || \chi^i - \Bar{\chi} ||
 $$
 \State Calculate all $\mathcal{N}^i$s at EOSIO block producer nodes ('\textit{nodeos}')
 \State Block producer nodes initiate a smart contract with blockchain accounts (\textit{'cleos create account'})
 \State \textbf{while} {$i \neq 0$}
\State \textbf{Contract Name:} SensorID\_timestamp (\textit{' cleos set contract'})
\State \textbf{Actions:}
\State \quad \textbf{addnovelty}
% \State \quad Add $\mathcal{N}$ to the table
 \State \quad \quad \textbf{Input:} account\_name, $\mathcal{N}^i$ 
\State \quad \quad \textbf{Require:} Authentication of user
% \State \quad \quad \textbf{Logic:} 
\State \quad \quad \quad Add $\mathcal{N}^i$ to the output table ('\textit{cleos push action}')
\State \quad \quad \quad Check if the NI is close to 1:
\begin{equation*}
\textbf{if } |\mathcal{N}^i - 1| < \epsilon \text{ then}
\end{equation*}
\State \quad \quad \quad \quad Display message: "Unhealthy bridge detected!"
\State \quad \quad \quad \textbf{else}
\State \quad \quad \quad \quad Continue 
\State \quad \quad \quad \textbf{end if}
 \State Consensus among EOSIO nodes
 \State Add $\mathcal{N}^i$ on the blockchain as input transactions
\end{algorithmic}
\end{algorithm}

\subsection{Smart Contract EOSIO using NI}
The first step in creating an EOSIO smart contract is writing the contract code in C++ which defines the actions, data structures, and logic that govern the behavior of the smart contract \cite{EOSIOxblock}. Once the smart contract code has been developed, it needs to be deployed to the EOSIO blockchain. Deployment involves uploading the compiled contract code to the blockchain network. This step requires an account with the necessary permissions to deploy contracts. Once deployed, the smart contract can be invoked by users or other smart contracts on the blockchain. Invoking a smart contract typically involves calling one of its actions, which triggers the execution of the corresponding code within the contract. When an action is invoked, the EOSIO blockchain executes the corresponding code in the smart contract. This code can read from and write to the blockchain's state, perform computations, and interact with other contracts and external data sources. During execution, the EOSIO blockchain validates the transaction and ensures that it meets the requirements specified by the smart contract code. Once the transaction is validated and executed, it is included in a block by the block producer nodes on the EOSIO blockchain. The block is then added to the blockchain, and the transaction becomes part of the immutable ledger. During execution, smart contracts can emit events to provide information about their state or the outcome of certain operations. These events can be logged and monitored by external systems or other smart contracts for tracking and auditing purposes. Fig. \ref{fig:sc-workflow} depicts the workflow of the EOSIO smart contract. EOSIO smart contract is called and deployed using the cleos commands. At the end of the execution of each smart contract, the novelty indices are added to the table, which is confirmed on the blockchain using nodeos. Nodeos is responsible for block producers, blockchain state, transaction validation, and consensus. The Algorithm \ref{alg:iot_blockchain} shows the process of collection of IoT data, calculation of NI, and confirming it onto the EOSIO blockchain using smart contract.

The algorithm, \ourmethod\  starts by collecting sensor response data and extracting relevant features from it using technique, SVD. These extracted features are stored for future reference, while the algorithm calculates the average of healthy feature vectors to establish a baseline representation of the bridge's normal behavior. Next, novelty indices are calculated for each dataset, measuring the deviation of the data from the established baseline. These novelty indices are then sent to EOSIO block producer nodes, where they are processed and added to the blockchain as input transactions. Within the blockchain environment, a smart contract is initiated with each nodeos producer, enabling the addition of novelty indices to the blockchain ledger. The algorithm also includes a logic to detect and respond to unhealthy bridge conditions by comparing the calculated novelty indices to a predetermined threshold value. If an NI is close to 1 then, an unhealthy bridge is detected, and a corresponding message indicating an unhealthy bridge is displayed. Finally, consensus among the EOSIO nodes is reached to confirm the validity of the added transactions, ensuring the integrity and security of the blockchain. 

\ourmethod\ algorithm offers an approach to bridge health monitoring by leveraging the capabilities of the EOSIO blockchain to facilitate secure and efficient data processing and unhealthy bridge detection.

\subsection{Data recovery from Blockchain}

Data recovery from the Eosio blockchain includes a list of cleos commands based on what kind of information needs to be retrieved. It is important to know the exact smart contract and table where the transaction came from and is confirmed on the blockchain. In \ourmethod\ we store each smart contract based on the bridge number and timestamp so that it is easy to track the sensors and retrieve the data. The command 'cleos get account' retrieves information about a specific account, 'cleos get transaction' retrieves information about a specific transaction, and 'cleos get block' retrieves information about a block.

% \subsection{Algorithm:\ourmethod}

\section{Performance Analysis}
In this section, we discuss the performance of \ourmethod\ in various scenarios. We begin by discussing the implementation details, followed by a thorough examination of the performance metrics.

\subsection{Implementation details}
% We collect real-time sensor data from a bridge in Nebraska State. The experimental data was collected on three days with different traffic loads. The bridge condition varies between healthy and unhealthy states.
% The transaction test performed throughout our research involved the iteration over each tdms file within a given folder, calculating its NI (NI), and storing the resulting values on a multi-index table on a blockchain following the EOSIO protocol. Our dataset contained 50 tdms files in total. Our experiments focused on the measurements produced by 51 strain sensors within the tdms files. Wanting to measure the performance of our blockchain over a longer period of time, our python script multiplied the array containing 50 novelty indices by 15. Therefore, 750 values were placed on the blockchain each time the transaction test was run. Analyzing log data created by the nodeos processes, we were able to plot and measure metrics such as throughput, resource consumption, latency, etc. To provide clarity, every test in our research 
% The experiments were conducted on a computer running Ubuntu 18.04 with 6169 MB of base memory and 8 CPU cores, with an AMD Ryzen 7 5800H processor clocked at 3.20 GHz,  16.0 GB RAM, and a 64-bit processor. For calculating the NI, we used Python 3.7, Pandas, Numpy, and Matplotlib. The smart contract was executed in CPP along with the EOSIO mainchain implementation.

Our experimental setup leverages real-time sensor data collected from a bridge in Nebraska State, covering multiple days with varying traffic loads and bridge conditions, including both healthy and unhealthy states. Executed on a computer running Ubuntu 18.04, equipped with 6169 MB of base memory, an AMD Ryzen 7 5800H processor, and 16.0 GB RAM, alongside a 64-bit processor architecture, our implementation utilizes Python 3.7, Pandas, Numpy, and Matplotlib for NI calculations. The integration of the NI within EOSIO smart contracts is facilitated by CPP within the EOSIO mainchain framework.

\begin{table}
\centering
% \scriptsize
\caption{Experimental Setup}
\label{tab:ppo_hprams}
\begin{tabular}{|l|l|}
\hline
\textbf{ Parameters }    & \textbf{Value }           \\ \hline
Total Sensors   & $51$     strain transducers                                           \\ \hline
% Sensor Types     & $3$   LVDT, $3$ accelorometer                                             \\ \hline
Maximum EOSIO nodes           & $50$                                               \\ \hline
Maximum smart contracts per node              & $50$                                                \\ \hline
Memory buffer size             & $150 MB$ 
     \\ \hline
Sampling rate         & $256$                                               \\ \hline
% Propagation delay from memory buffer        & $5ms$                                               \\ \hline
Data rate            & $100$Mbps                                                 \\ \hline
Propagation delay of data                       & 5ms                                               \\ \hline

\end{tabular}
\end{table}

\begin{figure}
    \centering
    \includegraphics[width= 0.9\linewidth]{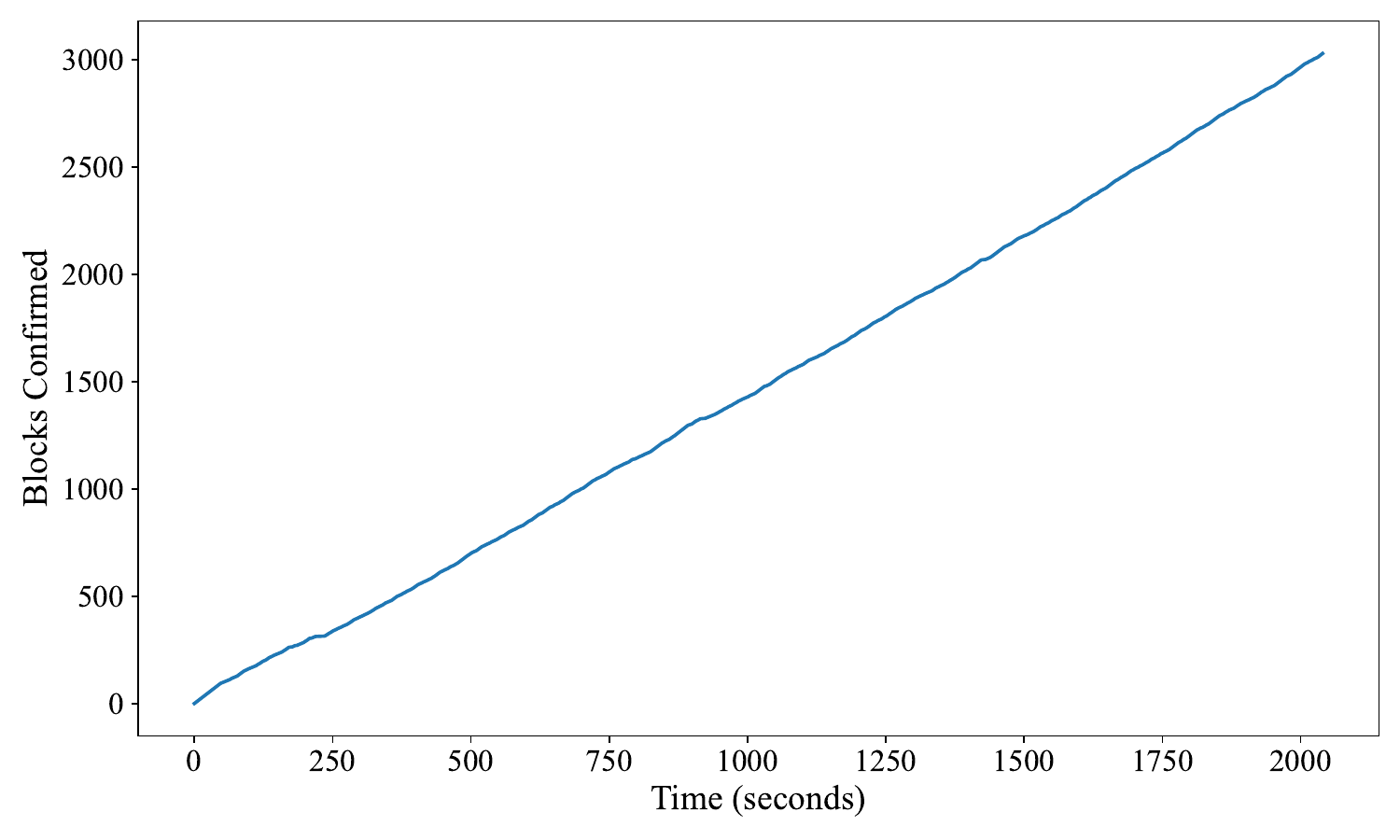}
    \caption{Performance of \ourmethod\ in terms of block confirmed per sec. A new block is confirmed every 0.5 sec.}
    \label{fig:block}
\end{figure}

\subsubsection{Comparison of with and without NI}
In \ourmethod\ we showed the implementation with NI on IoT data. The following table \ref{tab:NI_comp} summarizes the advantages of using NI instead of storing all the IoT data on the blockchain. Consider a blockchain with $N$ nodes, $k$ accounts created, and $s$ number of IoT sensors.  For every epoch of blockchain $m$ files from the memory buffer are processed by the smart contract. Each of the $m$ files has data of $i=1,2, \cdot, s$ sensors in its columns. A NI run on $m$ files gives $s$ number of NI values. Let $B$ be the size of the block. Assume that we know all the block details to retrieve data using cleos commands. For example, if we have 51 sensors and 500 files, we get 51 values NI. However, each file in 500 files has 50 data columns and 4396 rows entries making the total data 224196 values per file. Each of these values with or without NI is stored on the blockchain ledger in a distributed fashion. Data retrieval is performed using a multi-index table. This uses the find method on the table, which takes the primary key of the row that you want to retrieve and will return an iterator (reference) to the row. The size of multi index table created depends on the number of actions performed. To remove data from a table, the erase method is used which takes a reference to the iterator. Table \ref{tab:NI_comp} provides a comparison of executing \ourmethod\ with and without NI.

Comparing our method with and without the NI can provide valuable insights into the efficiency and effectiveness of our approach. In Table \ref{tab:NI_comp}, we can see the benefits of incorporating the NI, such as improvements in storage efficiency, data retrieval time, scalability with bridges, and latency per epoch. The proposed scheme significantly enhances storage efficiency by reducing the data stored on the blockchain. This optimization helps to improve the overall system performance by utilizing resources more effectively.

\begin{table}[]
    \centering
    \begin{tabular}{|c|c|c|}
    \hline
     Parameter    &  \ourmethod\ with NI & \ourmethod\ without NI\\ \hline
       Storage Efficiency  &  O(ns) & O(nms)\\ \hline
       Data Retrieval Time    & O(ks)  & O(kms)   \\ \hline
       Scalability with bridges   & O(n)  & O(n)   \\ \hline
       Latency per epoch & O(Bs) & O(Bms) \\ \hline
    \end{tabular}
    \caption{Comparision of \ourmethod\ with and without NI}
    \label{tab:NI_comp}
\end{table}

\begin{figure}
    \centering
    \includegraphics[width= 0.9\linewidth]{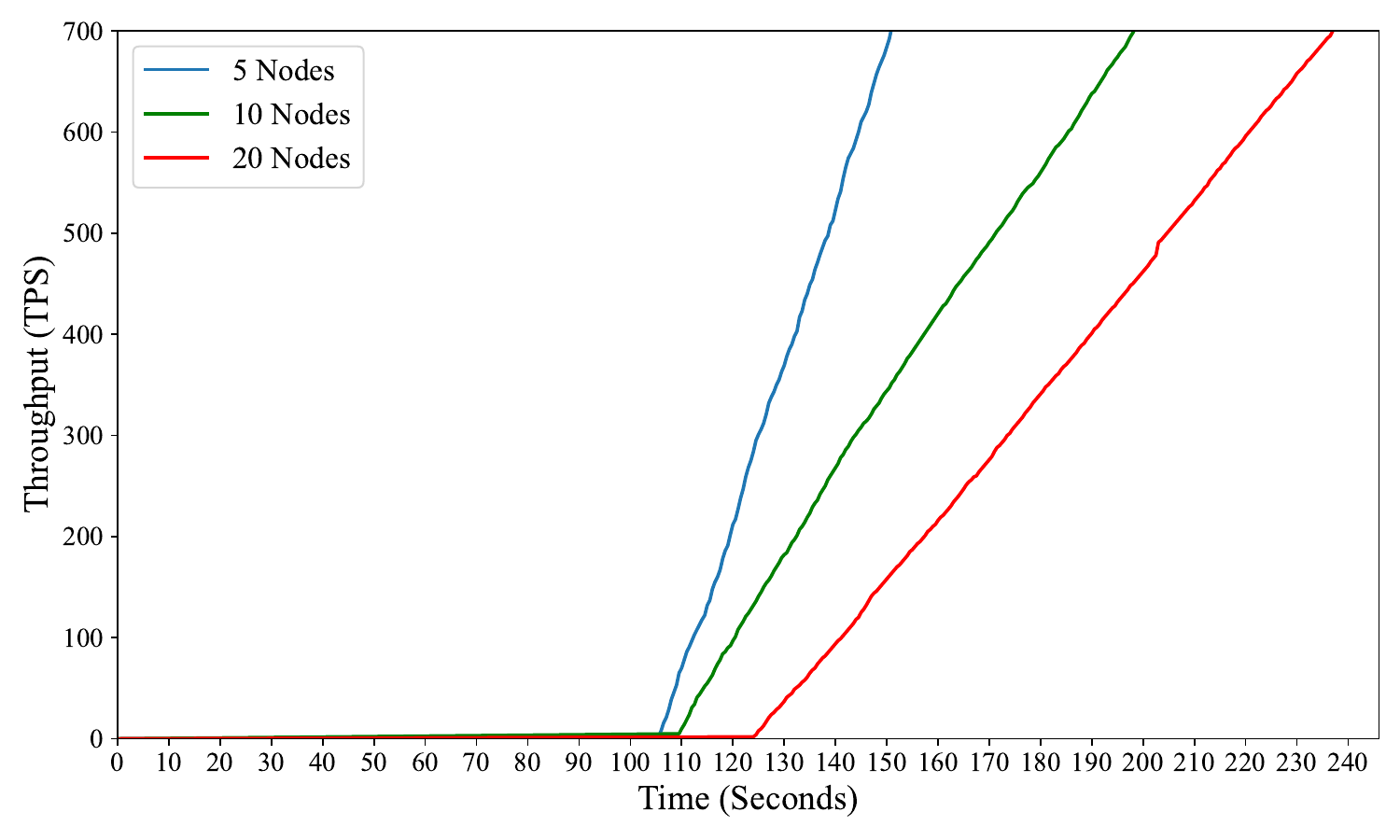}
    \caption{Performance of \ourmethod\ for increasing number of blockchain nodes. The transaction throughput increases proportionally with the number of nodes.}
    \label{fig:TPS_Plot_Num_Nodes}
\end{figure}

\begin{figure}
    \centering
    \includegraphics[width=0.8\linewidth, keepaspectratio]{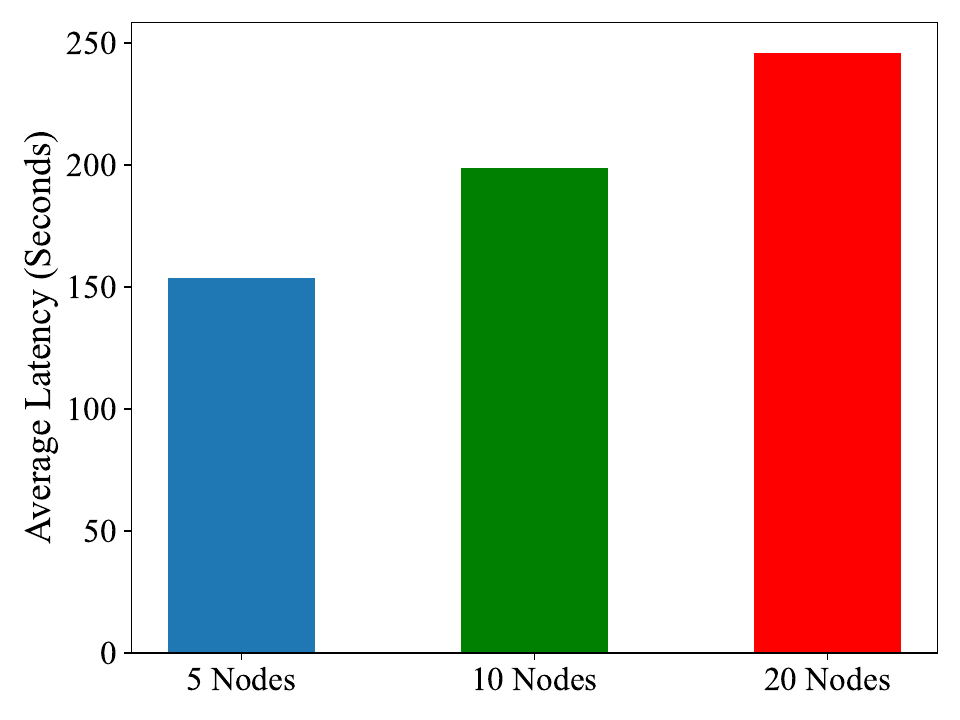}
    \caption{The average latency is plotted for 5,10, and 20 blockchain nodes. The latency increases with an increase in the number of nodes.}
    \label{fig:Latency_Chart_Num_Nodes}
\end{figure}

\begin{figure*}
    \captionsetup[subfigure]{}
    	\begin{center}
                \subfloat[\label{Cpu_5nodes}]{                			\includegraphics[width=0.45\linewidth,height= 4cm,keepaspectratio]{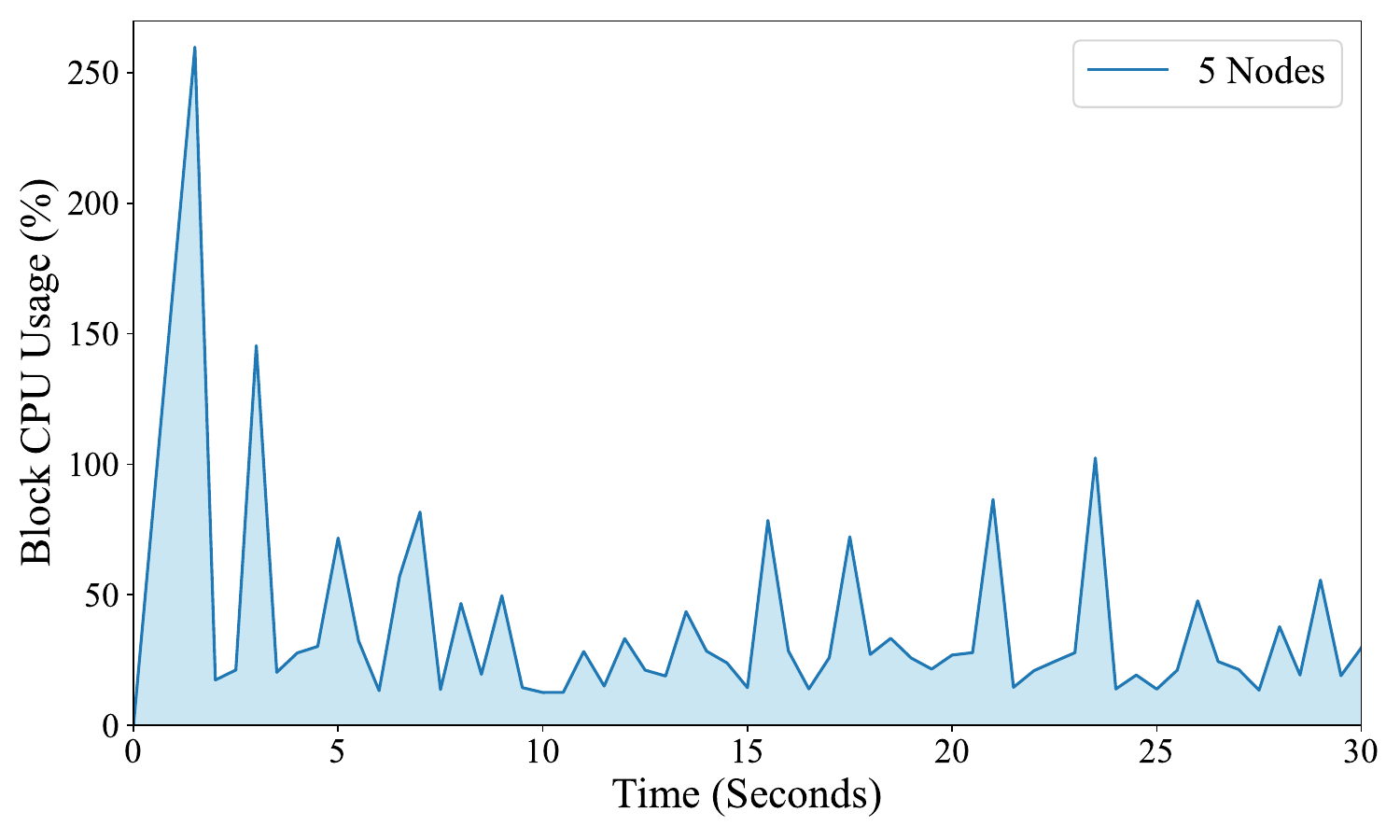}}
                \subfloat[\label{Cpu_10nodes}]{
                			\includegraphics[width=0.45\linewidth,height= 4cm,keepaspectratio]{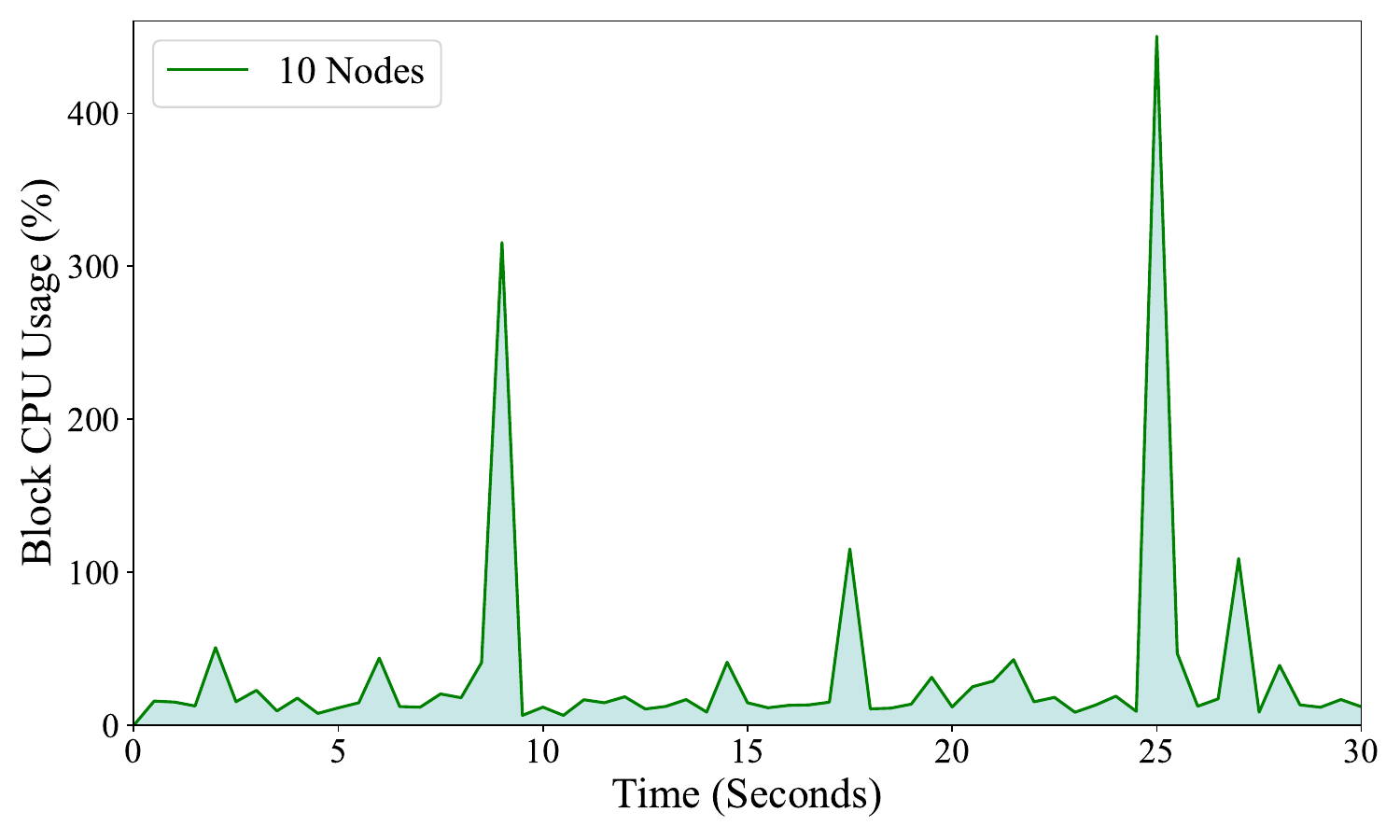}
                }\
                \subfloat[\label{Cpu_20nodes}]{
                			\includegraphics[width=0.45\linewidth,height= 4cm,keepaspectratio]{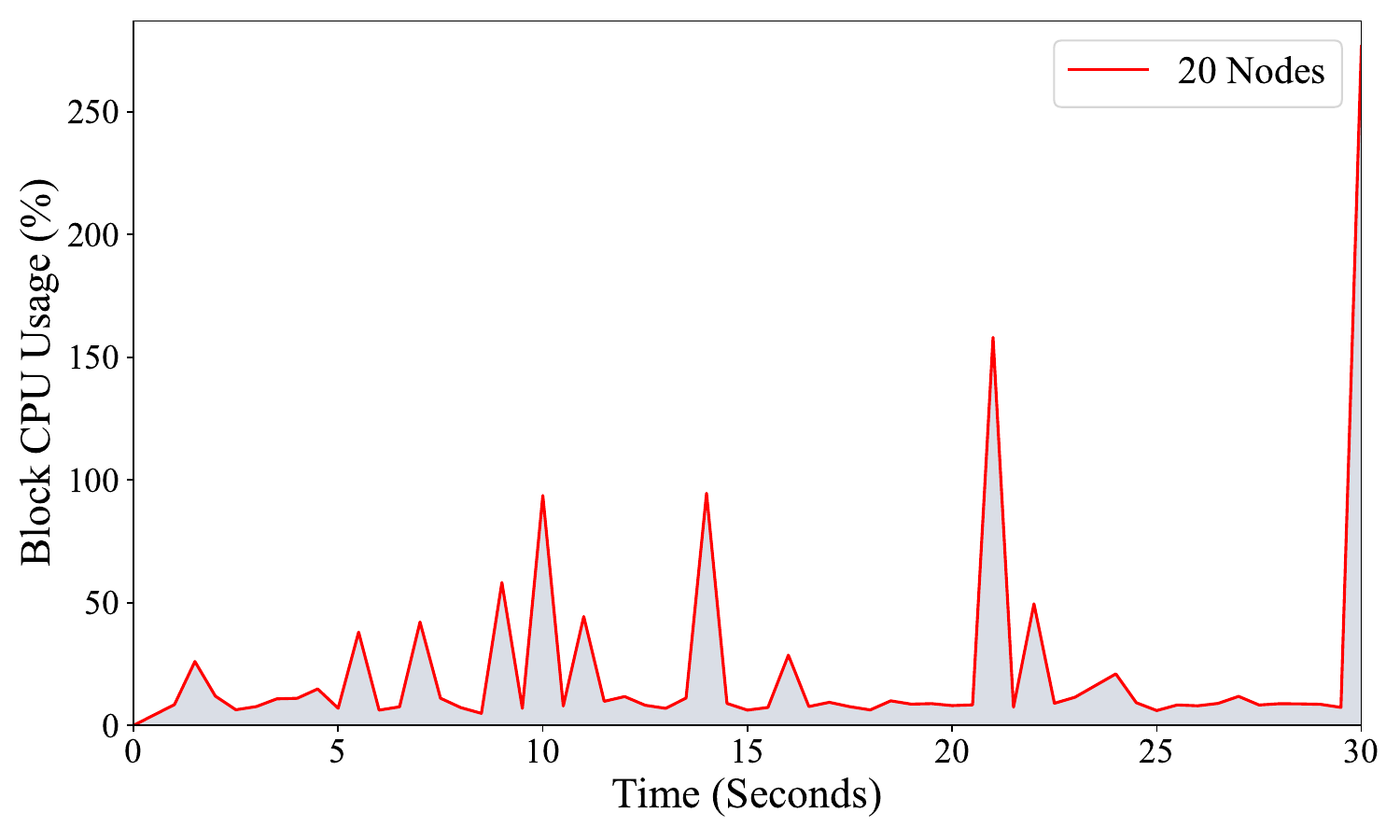}
                }
                \subfloat[\label{comp}]{
                			\includegraphics[width=0.45\linewidth,height= 4cm,keepaspectratio]{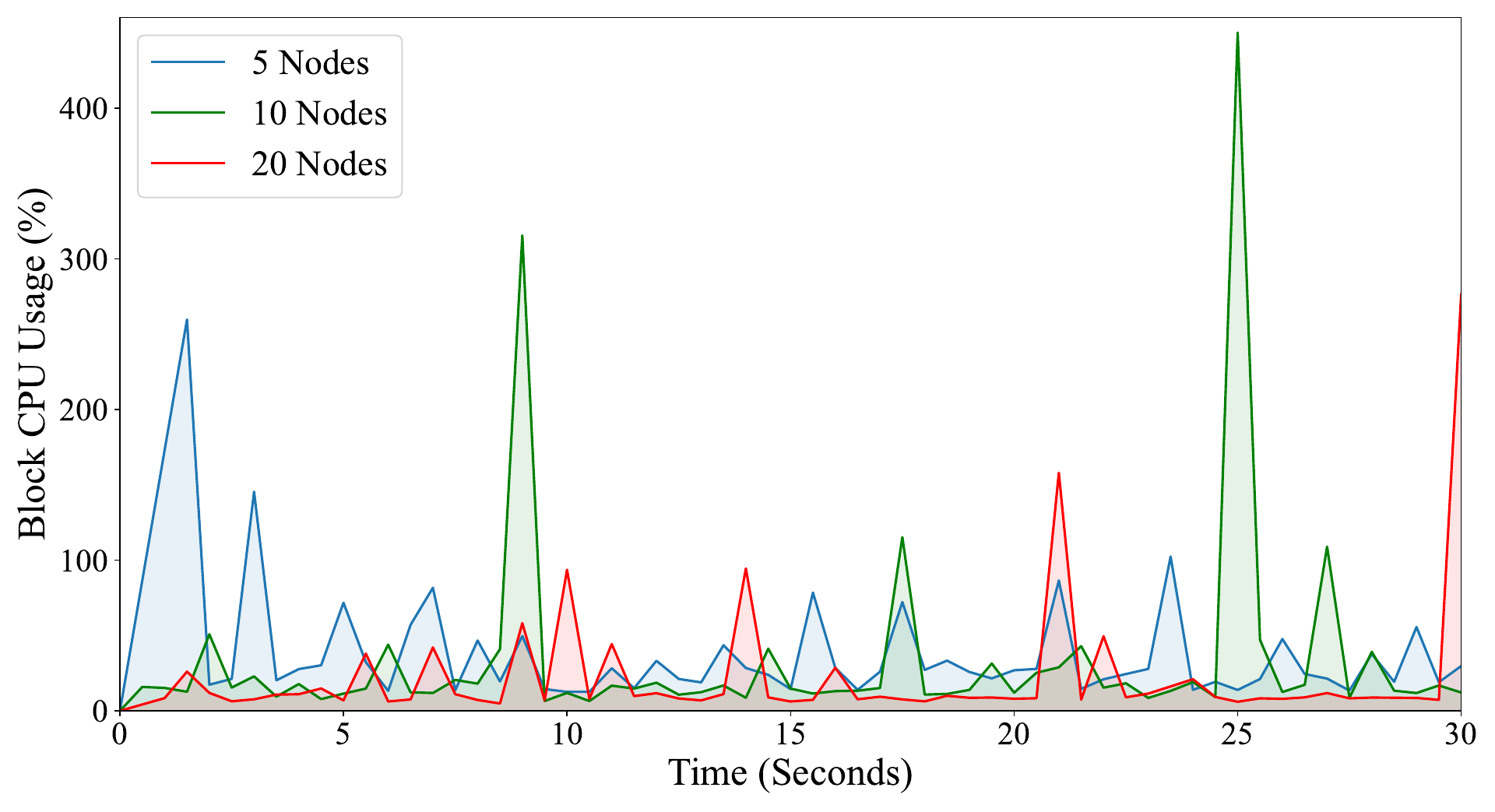} \label{fig:comp_cpu}
                }
        \end{center}
     \caption{The percentage of CPU usage for block confirmation is fairly consistent for 5, 10, and 20 blockchain nodes. - (a) 5 blockchain nodes (b)  10 blockchain nodes (c) 20 blockchain nodes (d) Comparison of block CPU usage for 5, 10, and 20 blockchain nodes.}
  \label{fig:CPU_Subplots_Num_Nodes}
\end{figure*}

% \subsection{Bridge Health Analysis with \ourmethod}

\begin{figure}
    \centering
    \includegraphics[width=0.9\linewidth,]{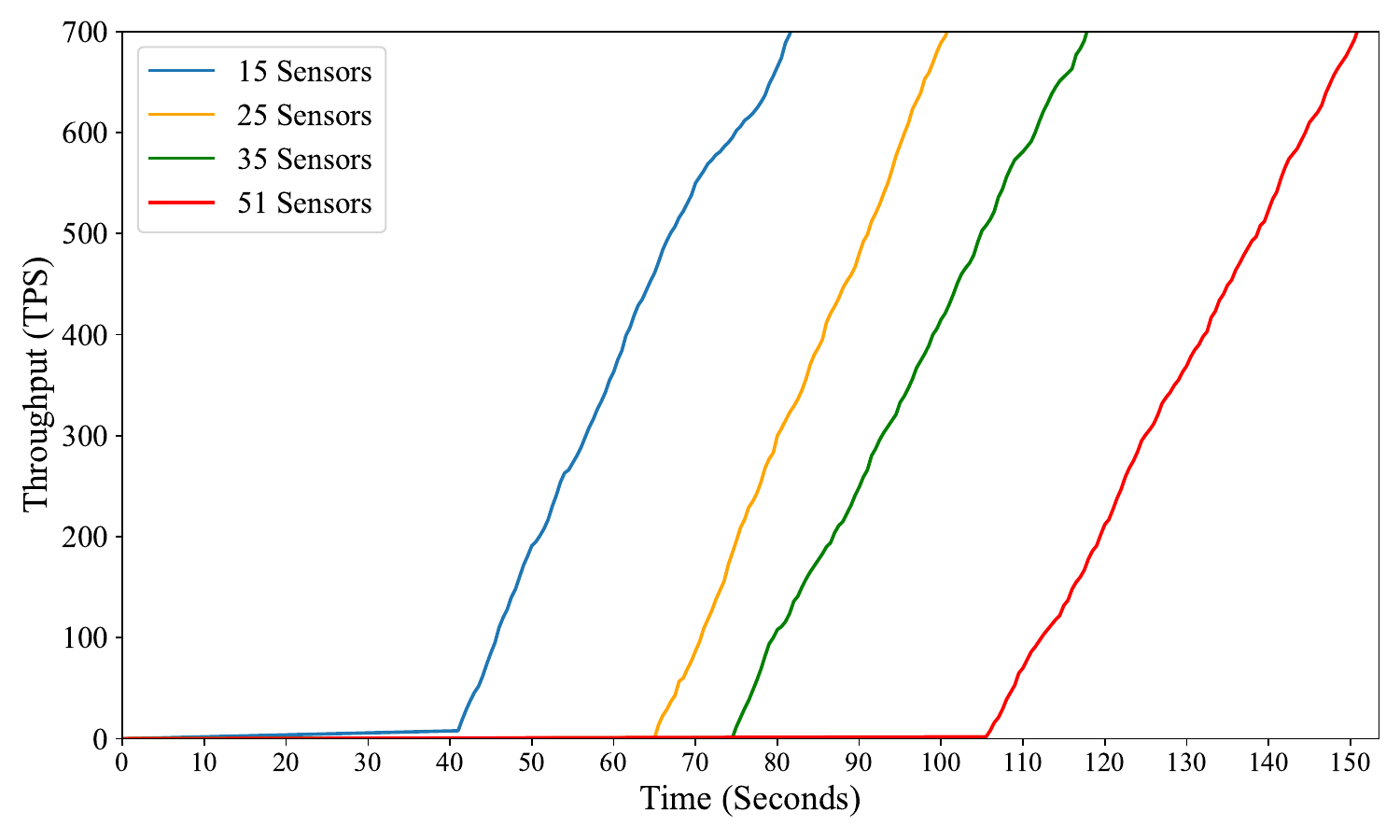}
    \caption{The throughput versus time of \ourmethod\ is plotted and is observed that the throughput scales with an increasing number of sensors.}
    \label{fig:TPS_Plot_Num_Sensors}
    \vspace{-0.8cm}
\end{figure}

\begin{figure*}[!htp]
    \captionsetup[subfigure]{}
    	\begin{center}
                \subfloat[\label{Cpu_15sensors}]{
                			\includegraphics[width=0.4\linewidth, height= 4cm, keepaspectratio]{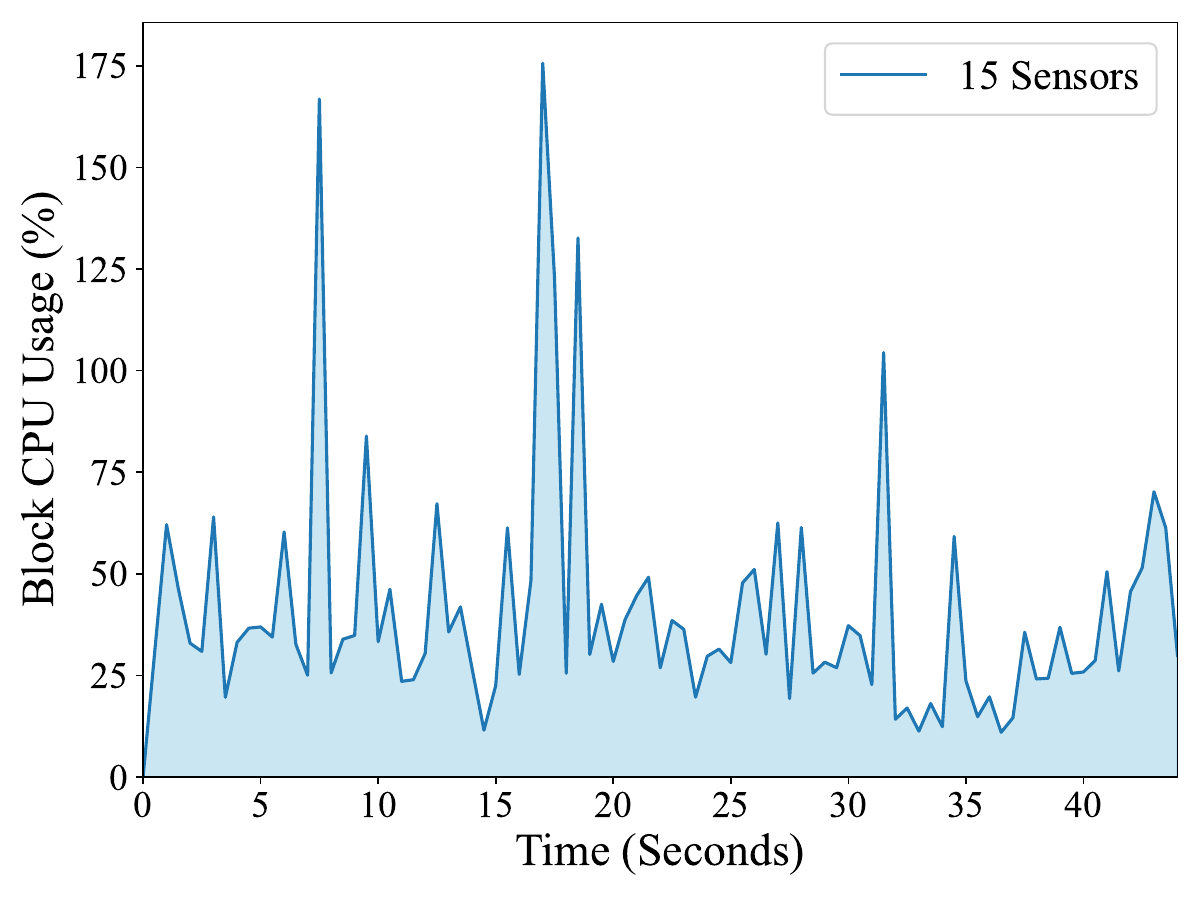}
                }
                \subfloat[\label{Cpu_25sensors}]{
                			\includegraphics[width=0.4\linewidth, height= 4cm, keepaspectratio]{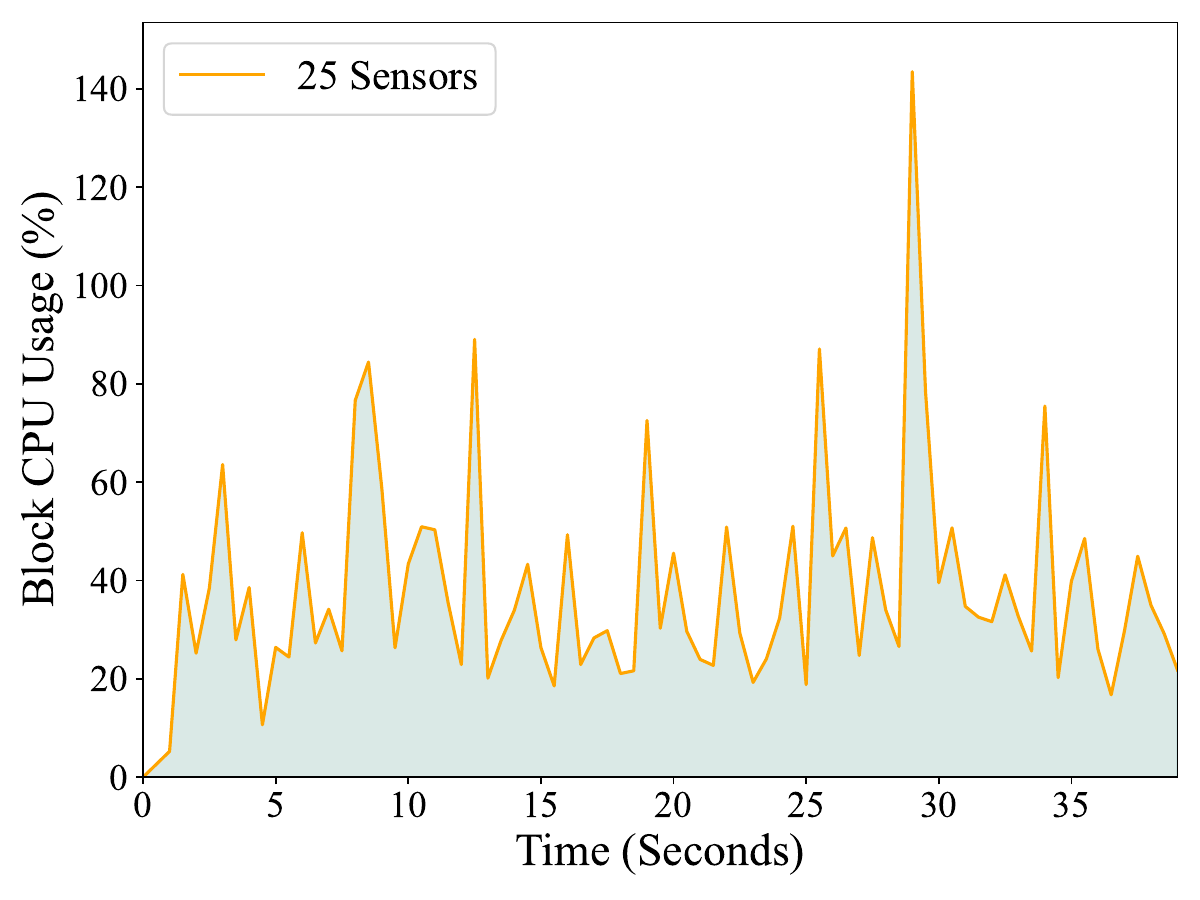}
                }\
                \subfloat[\label{Cpu_35sensors}]{
                			\includegraphics[width=0.4\linewidth,height= 4cm,keepaspectratio]{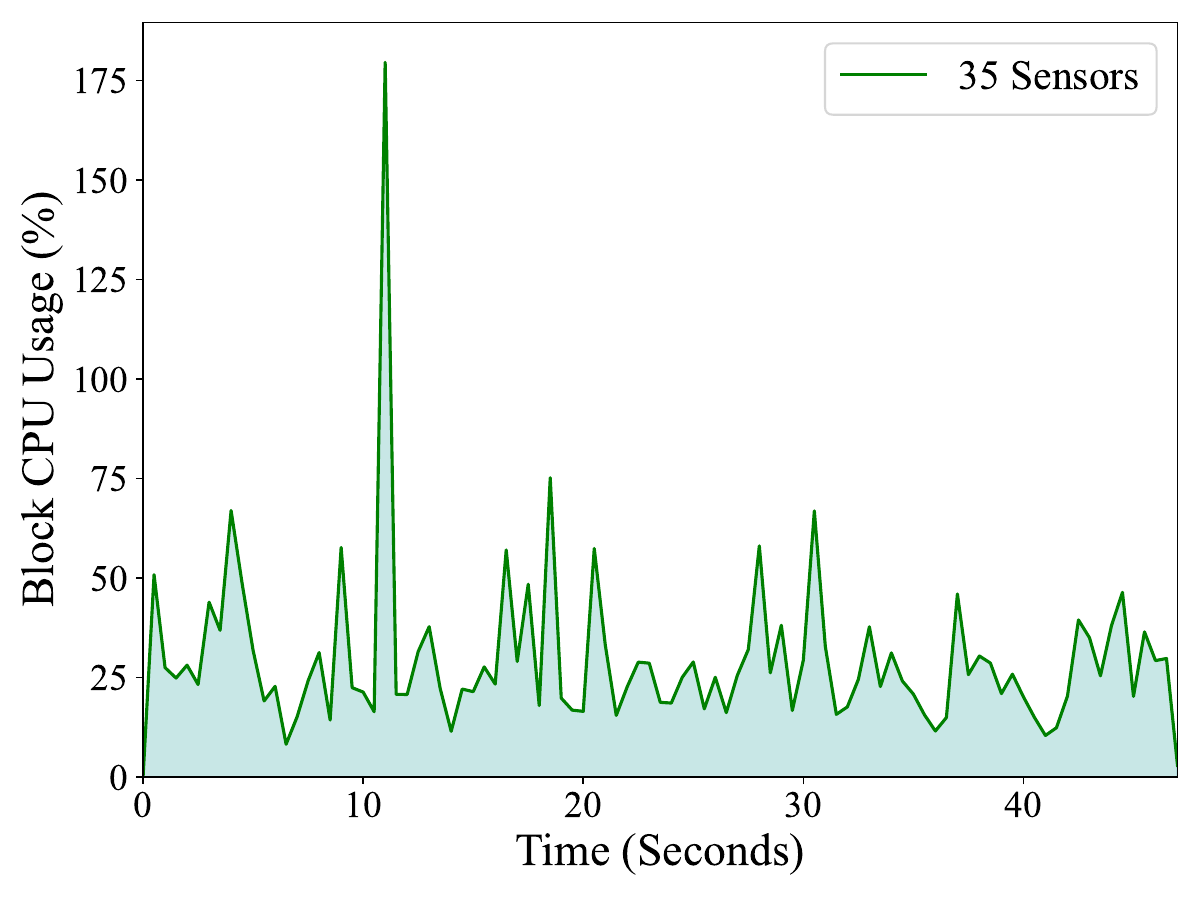}
                }
                \subfloat[\label{Cpu_51sensors}]{
                			\includegraphics[width=0.4\linewidth,height= 4cm,keepaspectratio]{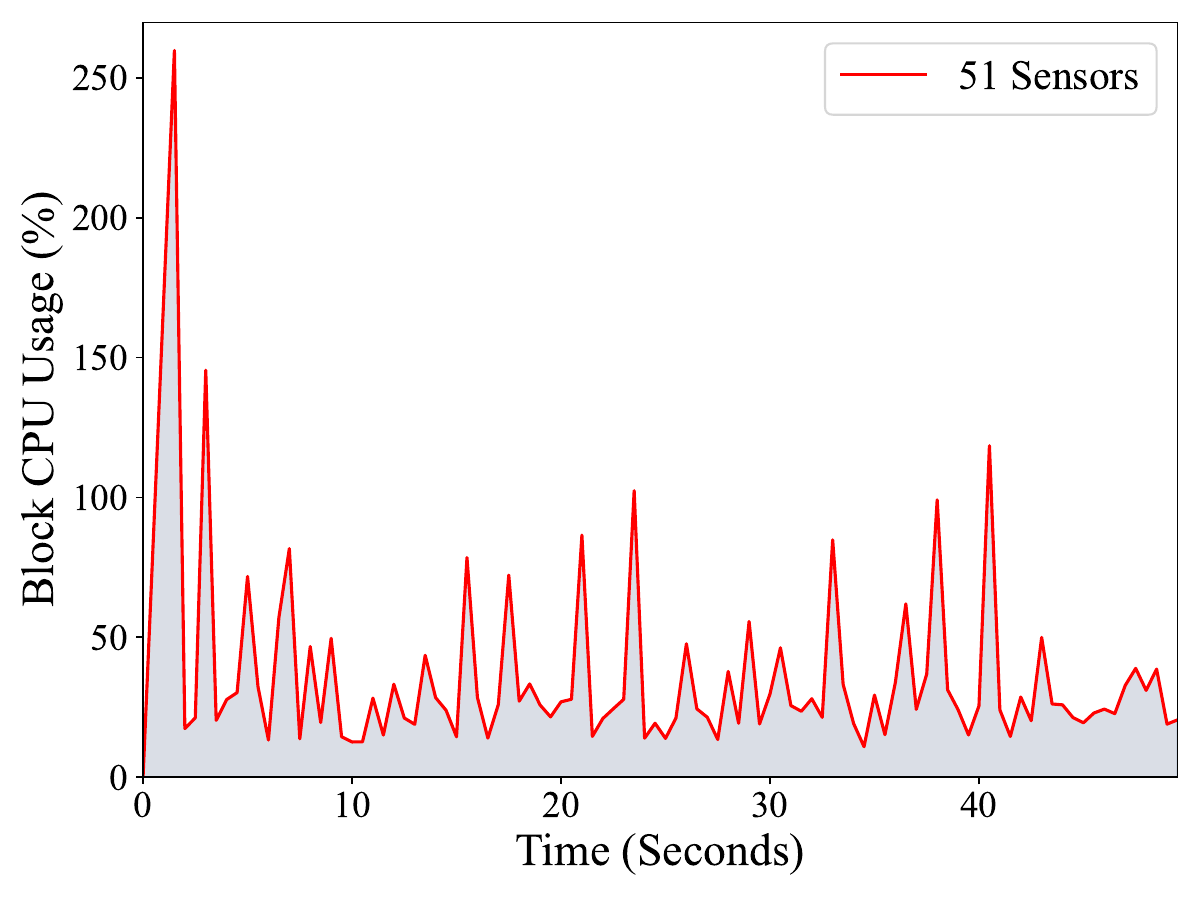}
                }
        \end{center}
     \caption{The percentage of CPU usage for block confirmation is fairly consistent for 15, 25, 35, and 51 bridge IoT sensors. - (a) 15 bridge IoT sensors (b)  25 bridge IoT sensors (c) 35 bridge IoT sensors (d) 51 bridge IoT sensors.}
  \label{fig:CPU_Subplots_Num_Sensors}
\end{figure*}

\subsection{Performance for increasing Blockchain Nodes}

In this section, we analyze the performance of \ourmethod\ as we increase the number of blockchain nodes. This investigation is crucial to understand how our approach scales as the network infrastructure expands. The EOSIO-based method used in \ourmethod\ confirms 2 blocks per sec. This is consistent throughout the experimentation and is depicted in Fig. \ref{fig:block}. The blocks confirmed by the blockchain linearly increase. Each block can consist of a varying number of transactions. Fig. \ref{fig:TPS_Plot_Num_Nodes} displays a trend of transaction throughput increasing proportionally with the number of nodes. This observation highlights the scalability of our approach, indicating that it can handle transactions efficiently even in larger network configurations.

Our analysis has revealed some interesting insights into the usage of resources. Fig. \ref{fig:CPU_Subplots_Num_Nodes} shows the block CPU usage as a percentage of 5000 us during the first 30 seconds of transaction execution. It is remarkable that the CPU usage remains stable despite an increasing number of blockchain nodes. This is demonstrated in Fig \ref{fig:comp_cpu}. This consistency in CPU usage can be attributed to the parallel processing capability of the EOSIO blockchain framework. By distributing computational tasks across multiple nodes, EOSIO ensures efficient resource utilization, enabling the CPU resources to handle transaction processing demands effectively.% It is important to take note that there are occasional spikes in CPU usage, especially during block confirmation at Layer 1 when the final block is added to the blockchain. These spikes are due to temporary increases in computational demand during intensive processing tasks, as shown in Fig. \ref{fig:CPU_Subplots_Num_Nodes}. Although these spikes may briefly strain system resources, they do not significantly affect the overall stability or performance of our approach.

In Fig. \ref{fig:Latency_Chart_Num_Nodes}, we observe the impact of increasing the number of blockchain nodes on the average latency of the system. As the number of nodes in the blockchain network grows, the process of reaching consensus becomes more complex and time-consuming. Consensus mechanism, DPoS requires all participating nodes to agree on the validity of transactions and the state of the blockchain. With more nodes involved, the consensus algorithm must coordinate a larger number of participants, leading to increased communication overhead and computational requirements. Consequently, the average latency, which measures the time taken for transactions to be confirmed and added to the blockchain, tends to rise as the network expands. This is depicted in Fig. \ref{fig:Latency_Chart_Num_Nodes}, where the average latency increases with the number of blockchain nodes. The higher latency observed with more nodes underscores the trade-off between decentralization and transaction speed inherent in blockchain networks. While a larger number of nodes enhances the network's resilience and security by distributing control among multiple parties, it also introduces delays in transaction processing due to the consensus process's increased complexity.
To enhance blockchain performance, it's important to know the link between node number and latency. Our study of \ourmethod\ with increasing nodes shows it's scalable and efficient. It can handle high transaction throughput and manage resources, making it suitable for large-scale deployments in bridge health monitoring applications. By managing resources, adjusting consensus parameters, and implementing scalability solutions, developers can mitigate latency effects caused by network growth, ensuring efficient transaction processing.

\subsection{Performance with increasing of IoT sensors}
In this section, we explore the impact of increasing the number of sensors on the performance of our proposed methodology, \ourmethod. This analysis is crucial for understanding how our approach manages changes in data volumes that arise with a growing number of sensors. This is a common scenario in bridge health monitoring applications.

We have observed a clear correlation between the number of sensors and the volume of transactions confirmed by our system, as shown in Fig. \ref{fig:TPS_Plot_Num_Sensors}. With each increase in the number of sensors, there is a corresponding rise in the number of transactions processed and confirmed by our method. This observation highlights the system's capability to handle and process sensor data efficiently, ensuring timely and reliable transaction confirmation even as the data volume escalates. The increasing trend in confirmed transactions demonstrates the scalability and robustness of our approach, proving its ability to adapt to changing data volumes without compromising performance. By systematically scaling the number of sensors, we gain valuable insights into the system's capacity to accommodate growing data loads, assuring its suitability for real-world deployment scenarios where sensor networks may expand over time. Moreover, the positive relationship between the number of sensors and confirmed transactions underscores the effectiveness of our approach in harnessing sensor data to derive meaningful insights and facilitate informed decision-making in bridge health monitoring. As the number of sensors increases, the volume of data available for analysis also increases, enabling more comprehensive and accurate assessments of bridge health and structural integrity.

In Fig. \ref{fig:CPU_Subplots_Num_Sensors}, we observe the CPU resource utilization in our system as the number of sensors is scaled. The plot shows a consistent pattern of CPU usage among different sensor configurations, highlighting the robust parallel processing capability of the EOSIO blockchain framework. The stable CPU usage demonstrates the efficiency of the EOSIO blockchain in distributing computational tasks across multiple nodes, ensuring optimal resource utilization even under fluctuating data volumes and network conditions.

% \begin{figure}
%     \centering
%     \includegraphics[width= 0.9\linewidth]{CPU_Subplots_Num_Sensors.pdf}
%     \caption{Caption}
%     \label{fig:CPU_Subplots_Num_Sensors}
% \end{figure}

% \begin{figure}
%     \centering
%     \includegraphics[width= 0.9\linewidth]{CPU_Plot_Num_Sensors.pdf}
%     \caption{}
%     \label{fig:}
% \end{figure}

\begin{figure}
    \centering
    \includegraphics[width= 0.8\linewidth]{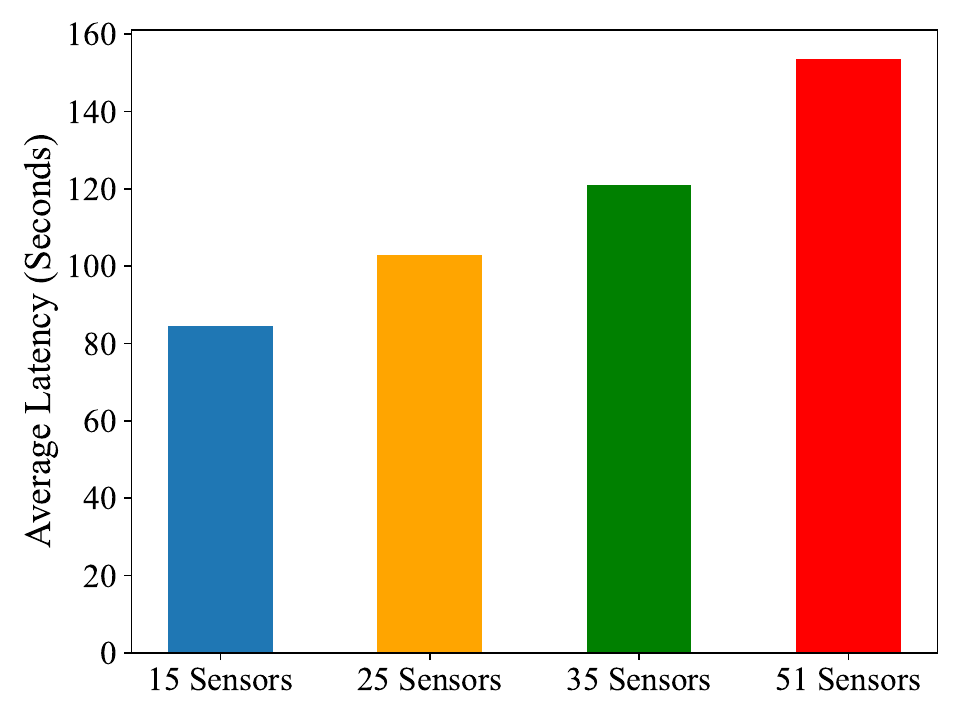}
    \caption{The average latency is plotted for  15, 25, 35, and 51 bridge IoT sensors. The latency increases with an increase in the IoT sensors.}
    \label{fig:Latency_Chart_Num_Sensors}
\end{figure}

This analysis examines the relationship between the number of sensors and the average latency of our system, as shown in Fig. \ref{fig:Latency_Chart_Num_Sensors}. The plot shows that as the number of sensors increases, the average latency also rises due to the larger volume of data generated by the growing sensor network. This increase in data volume adds computational overhead, resulting in longer transaction confirmation times. This is partly because of more time taken to calculate the NI with more sensor data. To accommodate increasing data volumes, it is crucial to optimize system resources and network architecture. The observed increase in latency highlights the scalability challenges associated with handling large-scale sensor deployments within decentralized systems. While the EOSIO blockchain framework provides robust parallel processing capabilities, increasing data volumes can increase latency.

\vspace{-0.3cm}

\subsection{Discussion}
We observe a scalability and latency tradeoff. As the number of sensors increases, the system's scalability is tested by its ability to handle growing data volumes and transaction throughput. However, this scalability may lead to increased latency as processing and validating larger amounts of data can result in longer transaction confirmation times. Another tradeoff exists between resource utilization and performance. The efficient resource utilization of the EOSIO blockchain framework is demonstrated by the consistent CPU usage observed across different sensor configurations. However, stable resource utilization levels may require sacrificing some performance metrics, such as transaction throughput or latency. A balance between maximizing system throughput and minimizing resource consumption needs to be achieved. A tradeoff also exists between data processing efficiency and data volume. An increase in the number of sensors results in a larger volume of data generated, which must be processed and validated by the system. The EOSIO blockchain framework offers robust parallel processing capabilities, but handling large data volumes may introduce latency issues.

In this work, we conducted experiments using EOSIO smart contracts integrated with a NI approach for analyzing bridge sensor data. Our experimentation was conducted within a localized setup. However, for scenarios involving larger-scale data and a higher number of bridges, we propose the utilization of the TELOS implementation of the EOSIO blockchain specifically tailored for IoT data management.

\section{Conclusion}
This research demonstrates how the EOSIO blockchain technology can be used to monitor the health of bridges. By utilizing data from IoT sensors, a technique called the NI is used to obtain valuable insights. This technique is based on clustering and helps to identify unusual patterns in the sensor data. The integration of smart contracts enhances security and enables predictions based on the NI. The proposed scheme, \ourmethod\, has the potential to securely store and analyze sensor data to enhance bridge health monitoring systems. The effectiveness of the scheme is evaluated by analyzing real-world IoT sensor data from healthy and unhealthy bridges. The evaluation also includes scenarios with varying numbers of sensors. This work contributes to the growing field of blockchain applications, particularly in the infrastructure monitoring context.

In the future, this approach can be extended to multiple bridges throughout the state by leveraging the virtual private blockchain feature of the EOSIO blockchain. The researchers also plan to conduct further analysis beyond NI calculation for bridge health monitoring.

\bibliographystyle{IEEEtran}
\bibliography{IEEEabrv,reference.bib}

\end{document}